\documentclass[author,usenatbib,usecmfonts,reqno]{nrc2}

\usepackage[french,english]{babel}
\usepackage{cite}
\usepackage{bm}
\usepackage{amssymb}
\usepackage{amsmath}
\usepackage[figuresright]{rotating}
\usepackage{graphicx}
\usepackage[usenames,dvipsnames]{color}
\usepackage{lipsum}
\usepackage{ulem}
\usepackage[usenames,dvipsnames]{color}

\newcommand{\pp}[1]{\phi^{#1}}
\newcommand{\ff}[2]{\left(\nabla\phi\right)^{#2}_\mathrm{#1}}
\renewcommand{\d}{\mathrm{d}}

\newcommand{\beq}{\begin{equation}}
\newcommand{\eeq}{\end{equation}}
\newcommand{\bey}{\begin{eqnarray}}
\newcommand{\eey}{\end{eqnarray}}

\newcommand{\Msun}{\ensuremath{\text{M}_\odot}}

\newcommand{\PhiN}{\phi}
\newcommand{\Phiph}{\Phi_\text{ph}}
\renewcommand{\d}{{\text{d}}}
\newcommand{\grad}{{\bf \nabla}}

\newcommand{\ttt}[1]{\ensuremath{\times 10^{#1}}} 
\newcommand{\unit}[1]{\ensuremath{\,\text{#1}}}

\newcommand{\mnras}{MNRAS}

\newcommand{\apj}{ApJ}
\newcommand{\apjl}{ApJL}
\newcommand{\apjs}{ApJS}
\newcommand{\aap}{A\&A}
\newcommand{\nat}{Nature}

\newcommand{\prd}{Phys. Rev. D}

\newcommand{\aj}{AJ}

\newcommand{\aapr}{AAPR}
\newcommand{\amp}{AMP}
\newcommand{\RAMSES}{R{\sc amses}\xspace}
\newcommand{\POR}{P{\sc or}\xspace}

\newcommand{\NMODY}{{N-\sc mody}\xspace}
\newcommand{\NEMO}{N{\sc emo}\xspace}

\author{Fabian L\"ughausen}
\correspond{fabian@astro.uni-bonn.de}
\address[label1]{Helmholtz-Institut f\"ur Strahlen- und Kernphysik, Nussallee 14--16, D-53115 Bonn, Germany}
\author{Benoit Famaey}
\address[label2]{Observatoire astronomique de Strasbourg, Universit\'e de Strasbourg, CNRS UMR 7550, 11 rue de lÕUniversit\'e, F-67000 Strasbourg, France}
\author{Pavel Kroupa}
\address[label1]

\shortauthor{F.~L\"ughausen, B.~Famaey, and P.~Kroupa}

\begin{document}

\title{Phantom of RAMSES (POR): A new Milgromian dynamics N-body code}

\begin{abstract} 
Since its first formulation in 1983, Milgromian dynamics (MOND) has been very successful in predicting the gravitational potential of galaxies from the distribution of baryons alone, including general scaling relations and detailed rotation curves of large statistical samples of individual galaxies covering a large range of masses and sizes. Most predictions however rely on static models, and only a handful of $N$-body codes have been developed over the years to investigate the consequences of the Milgromian framework for the dynamics of complex evolving dynamical systems. In this work, we present a new Milgromian $N$-body code, which is a customized version of the \RAMSES{} code (Teyssier 2002) and thus comes with all its features: it includes particles and gas dynamics, and importantly allows for high spatial resolution of complex systems due to the adaptive mesh refinement (AMR) technique. 
It further allows the direct comparison between Milgromian simulations and standard Newtonian simulations with dark matter particles.
We provide basic tests of this customized code and demonstrate its performance by presenting $N$-body computations of dark-matter-free spherical equilibrium models as well as dark-matter-free disk galaxies in Milgromian dynamics.
\end{abstract}

\maketitle

\section{Introduction} 

The large-scale structure of the Universe reveals a dark sector generally supposed to be made of dark energy (e.g.\cite{Komatsu2011,Planck2013arXiv}) and hitherto undetected missing matter or dark matter (DM), the physical nature of both still being unexplained to date. In the current standard model of cosmology, DM is assumed to be made of non-baryonic particles, the so-called `cold dark matter' (CDM). Whilst successful on large scales, this interpretation leads to a long list of problems on the scales of galaxies and the Local Group \cite{Kroupa1,Kroupa3,Ibata2013Andromeda,Ibata2014,Pawlowski14prep,Bournaud2007,Gentile2007}. One of the biggest challenges for the DM-based model may be the tight scaling relations which galaxies follow \cite{FamMcgaugh}, the tightness of which cannot be unterstood in the DM context. These relations can all be well summarized by Milgrom's formula, devised more than 30 years ago \cite{Mil83},
\begin{equation}
\label{eq:milgromsformula}
	g = \left( g_\text{N} a_0 \right)^{1/2} \; {\rm for} \, g \ll a_0 \simeq 10^{-10} {\rm m}{\rm s}^{-2} \,.
\end{equation}
This is equivalent to stating that, for $g \ll a_0$, (i) the dynamics becomes scale-invariant under space-time transformations 
$(t,\boldsymbol x) \rightarrow (\lambda t,\lambda \boldsymbol x)$ with $\lambda \in \mathbb{R}$
\cite{Milgrom2009}, or (ii) gravity is anti-screened with a `gravitational permittivity' equal to $g/a0$ in some gravitationally polarizable medium \cite{BlanchetBernard2014arxiv}. On cosmological scales the formula does not work, meaning that Milgromian dynamics is only an effective description of dynamics on galaxy scales, and is probably part of a yet incomplete paradigm. But on galaxy scales, the formula is extremely succesful: not only were galaxy scaling relations correctly predicted by this formula (e.g., the baryonic Tully-Fisher relation \cite{McGaugh2005,StacyPRL}), some were actually found because they were pointed to by the formula (e.g., the mass discrepancy-acceleration relation \cite{Sanders1990,Scarpa2006,McGaugh2012}). So, independently of its deepest physical meaning, Milgrom's formula already achieved one of the most important roles of a theoretical idea, i.e. to direct posterior experiments or observations, and correctly predict their outcome. For a complete and up-to-date review of Milgromian dynamics, see \cite{FamMcgaugh}.

However, applying blindly Eq.~\ref{eq:milgromsformula} into a $N$-body code would lead to dramatically unphysical predictions. 
It can readily be seen that, in a two-body configuration, the force is not symmetric in the two masses. This means that Newton's third law (the action and reaction principle) does not hold in this framework, and as a consequence the momentum is not conserved. Instead of Eq.~\ref{eq:milgromsformula}, one thus has to use a non-linear generalization of Poisson's equation leading precisely to Eq.~\ref{eq:milgromsformula} in one-dimensional-symmetric configurations such as spherical symmetry. Two such classical modifications of the Poisson equation have been proposed \cite{BM84,QUMOND}, and have subsequently been shown to be natural weak-field limits of relativistic modified gravity theories \cite{TeVeS,BIMOND}. Because these modified Poisson equations all necessarily involve at least one non-linear step, the total force on each particle cannot be obtained by summing the individual forces from all the other particles as in Newtonian gravity. This means that one must rely on particle-mesh techniques to develop Milgromian $N$-body codes.

Only a handful of such Milgromian codes have been developed over the years. This field was pioneered by \cite{BradaMilgrom1999}, who studied the stability of galaxy disks, the formation of warps and the dynamics of satellite galaxies orbiting a host (gas dynamics was not considered). More recent codes, like \NMODY{} \cite{nmody}, solve the Milgromian Poisson equation in spherical coordinates and are therefore predestinated for close to spherically symmetric problems, but do not allow generic simulations of flattened and multicentered stellar systems on galactic and sub-galactic scales. Other codes are rather designed only for cosmological simulations \cite{LlinaresKnebe}. The most precise $N$-body computations for disk galaxies have been made by \cite{Tiret2007,Tiret2008}, but only with rudimentary treatment of hydrodynamics and very low resolution. The most recent code has been developed by \cite{Angus2,Angus2014}, based for the first time on the modified Poisson equation of \cite{QUMOND}.

To date, with so few codes and simulations at hand, only very little is actually known about the time-evolution of dynamical objects within Milgromian dynamics. Generic, fully dynamical tests using $N$-body codes with live particles and a full treatment of hydrodynamics are still missing. Testing Milgromian dynamics in dynamical $N$-body systems is however fundamental, because the implications of Milgrom's force law on the time-evolution of dynamical systems are not trivial, and analytic approaches with static realizations of galactic potentials are not sufficient to understand those implications. Even though the Milgromian force law mimics the potential of an effective DM halo, the physical properties of this effective DM halo are very different from those of the classical, pressure-supported DM halos, because it is, contrary to CDM halos, directly connected to the distribution of baryonic matter \footnote{This applies particularly to the formation of structure and substructure (e.g. the instability of rotating disk as in Sect.~\ref{sect:applications}).}, and is affected by non-trivial phenomena such as the so-called external field effect. What is more, this effective halo does not lead to any kind of dynamical friction.

Therefore, it is still unclear whether the Milgromian framework can successfully explain all dynamical probes of the gravitational potential on galactic scales, such as the geometry of tidal streams or even a vast topic like the formation of galaxies in general. Here we present a new Milgromian dynamics $N$-body code, which is a customized version of the \RAMSES{} code \cite{Teyssier2002} based on the modified Poisson equation of \cite{QUMOND}.  It comes with all its features, including adaptative mesh refinement, particle and gas dynamics, and it is suited for various different contexts such as the simulations of isolated systems, interacting systems, as well as the formation of structures in a cosmological context.

\section{Milgromian dynamics and QUMOND}
Since its first formulation in 1983 \cite{Mil83}, many relativistic and non-relativistic Milgromian gravity theories have been developed, yielding the scale-invariant property of Milgrom's empirical formula (Eq.~\ref{eq:milgromsformula}) in the weak-field limit and in spherical symmetry.
One such theory is the so called quasi-linear formulation of MOND (QUMOND) \cite{QUMOND}, having the following Poisson equation:
\begin{equation}
\nabla^2 \Phi (\bm x)
= 4 \pi G \rho_\text{b}(\bm x) + 
\nabla \cdot \left[ \nu\left(|{\mathbf\grad} \PhiN|/a_0\right) {\mathbf\grad} \PhiN(\bm x) \right]
\label{eq:poisson}
\end{equation}
or
\begin{equation}
\nabla^2 \Phi (\bm x)
= 4 \pi G \left( \rho_\text{b}(\boldsymbol x) + \rho_\text{ph}(\boldsymbol x) \right) \,.
\label{eq:poisson2}
\end{equation}
In this equation, 
$\rho_\text{b}(\bm x)$ is the baryonic density, 
and $\PhiN(\bm x)$ the Newtonian potential such that $\nabla^2 \PhiN(\bm x)=4 \pi G \rho_\text{b}(\bm x)$.
The function $\nu(y)$ is defined by the limits
 $\nu(y) \rightarrow 0$ if $y \gg 1$ (Newtonian regime) 
 and $\nu(y) \rightarrow y^{-1/2}$ if $y \ll 1$ (Milgromian regime).
We write the second term on the right hand side of Eq.~\ref{eq:poisson} as
\begin{equation}
	\rho_\text{ph}(\bm x) = \frac{\nabla \cdot \left[ \nu\left(|{\mathbf\grad} \PhiN(\bm x)|/a_0\right) {\mathbf\grad} \PhiN(\bm x) \right]}{4\pi G}
\label{eq:rho_phantom}
\end{equation}
in order to emphasize the quasi-linearity of this Poisson equation. 
It tells us that the Milgromian gravitational potential, $\Phi$, is defined by the baryonic matter density distribution, $\rho_\text{b}$, plus one additional term, noted as $\rho_\text{ph}$, which also has the unit of matter density and which is defined by the Newtonian potential, $\phi$, i.e. by the distribution of baryonic matter through Eq.~\ref{eq:rho_phantom}.
In other words, the total gravitational potential, $\Phi = \PhiN + \Phiph$, can be divided into two parts: a classical (Newtonian) part, $\PhiN$, and an additional Milgromian part, $\Phiph$.

The additional matter density distribution, $\rho_\text{ph}(\bm x)$, that would, in Newtonian gravity, yield the additional potential $\Phiph(\bm x)$, and therefore obeys $\nabla^2 \Phiph(\bm x) = 4\pi G \rho_\text{ph}(\bm x)$, is known as the `phantom dark matter' (PDM) density (Eq.~\ref{eq:rho_phantom}).
This PDM is not real matter but a mathematical ansatz that allows to compute the additional gravitational potential predicted by the Milgromian framework. Furthermore, it gives it an analogue in Newtonian dynamics and allows a convenient comparison of the predictions of Milgromian dynamics to those of the CDM-based standard model: the mathematical PDM density is exactly the density that would be interpreted as unseen  or dark matter in the context of the standard model of cosmology.

In the following Sect~\ref{sect:poissonsolver}, we explain in three simple steps how Eq.~\ref{eq:poisson} can be solved in general.
In Sect.~\ref{sect:PDM}, we exhibit a numerical scheme that allows to evaluate Eq.~\ref{eq:rho_phantom} on a diskrete grid.

\section{The `Phantom of RAMSES' (POR) code}
The goal of any $N$-body code is the computation of the force acting on the individual particles in order to integrate these particles through phase-space. In the simplest case, that is a system made of $N$ particles with known masses, positions and velocities, $N(N-1)$ forces have to be evaluated (if no simplifications are made). With modern computer technology, this is feasable for systems with $N \lesssim 10^{6}$, but computationally too expensive for systems with $N \gtrsim 10^6$.
However, in collisionless systems, i.e. systems which have relaxation time-scales longer than one Hubble time (this is by definition the case for galaxies), the potential is smooth and direct $N$-body forces can be neglected. 
This allows us to make a variety of simplifications to compute the accelerations for a large number of particles.
For instance tree-codes use a hierarchical spatial tree to define localized groups of particles whose contribution to the local gravitational field is calculated all at once. However, due to the non-linearity of Milgromian dynamics, this type of technique is in principle difficult to apply here. 
So the general approach used here is to map the particles (positions and masses) on a diskrete grid in order to determine the smoothed matter density distribution $\rho_\mathrm{b}(\boldsymbol x)$ of the particles.
From that, the gravitational potential $\phi(\boldsymbol x)$ is derived by solving the Poisson equation, 
\begin{equation}
	\nabla^2 \phi(\boldsymbol x) = 4\pi G \rho_\mathrm{b}(\boldsymbol x) \,.
\end{equation}
From this grid-based potential, the acceleration $\boldsymbol a(\boldsymbol x) = -\nabla\phi(\boldsymbol x)$ can be computed.
The acceleration is then interpolated at the positions of the individual particles, and the equations of motion for each particle are integrated.
These integrals can be approximated numerically in various ways. 

The \POR{} code is a customization of the \RAMSES{} code \cite{Teyssier2002} and therefore comes with all its features.
\RAMSES{} implements an adaptive mesh refinement (AMR) strategy \cite{Kravtsov97}. Starting with a coarse Cartesian grid, the grid is recursively refined on a cell-by-cell and level-by-level basis. If a cell fulfills certain refinement-criteria, e.g. exceeds a minimum particle number density, the cell is split up into $2^3$ sub-cells. This technique allows to use large simulation boxes and still to achieve high spatial resolution in the regions of interest. This makes the code particularly interesting not only for cosmological simulations, but also for e.g. galaxy-galaxy interactions, satellite galaxies in the field of a host galaxy, and more.
The Poisson equation is solved using a multi-grid relaxation scheme \cite{Kravtsov97,GuilletTeyssier2011}: the residual $\nabla^2 \phi - 4\pi G \rho$ is minimized iteratively using the Gauss-Seidel method. This approach has the advantage of allowing arbitrarily shaped domain boundaries and can therefore save a lot of work overhead. What is more, the convergence rate does not depend on the number of cells. This makes it particularly interesting for runs with a large number of particles/cells.
Once the Poisson equation is solved and the gravitational acceleration of each particle is known, a predictor-corrector scheme with adaptive time steps is used to move the particles through phase-space.

To solve Eq.~\ref{eq:poisson} on the adaptive grid, the Poisson solver has been patched to implement the procedure described in the following subsection.
The user can decide whether the Newtonian or the Milgromian force should be computed and applied by setting a flag in the project's configuration file without the need of recompiling the code. Having a code that can run Newtonian (\RAMSES{}) and Milgromian (\POR{}) $N$-body simulations at the same time is important, because it allows a side-by-side comparison of Newton$+$DM on one hand and Milgromian dynamics (without DM) on the other hand. We expect from this direct comparison to find qualitative differences between both scenarios leading to observationally testable predictions that can clearly distinguish between both cases.

\subsection{The modified Poisson solver}
\label{sect:poissonsolver}

Solving the QUMOND Poisson equation (Eq.~\ref{eq:poisson}) requires solving two linear differential equations and one additional algrebraic step.
\begin{enumerate}
\item
As before, the classical Poisson equation,
\begin{align}
	\nabla^2 \phi(\boldsymbol x) = 4\pi G \rho_\mathrm{b}(\boldsymbol x) \,,
\end{align}
is solved to compute the Newtonian potential, $\phi$, and its gradient, $\nabla\phi$, from the given matter density distribution defined by the particles and the gas. At the coarse grid boundary, the condition $\phi(r) = G M_\text{b} / r$ is applied. $M_\text{b}$ is the total baryonic mass, and $r$ the distance to its center.\footnote{The simulation box has to be chosen large enough to fulfill this condition at the coarsest level boundary.}
\item 
The PDM density $\rho_\text{ph}(\boldsymbol x)$ (Eq.~\ref{eq:rho_phantom}) is calculated applying the diskrete scheme which is described in detail in Sect.~\ref{sect:PDM}.
\item
Then, the Poisson equation is solved a second time, now with the baryonic matter plus PDM density,
\begin{align}
	\nabla^2 \Phi(\boldsymbol x) = 4\pi G \left( \rho_\mathrm{b}(\boldsymbol x) + \rho_\mathrm{ph}(\boldsymbol x) \right) \,,
\end{align}
to obtain the Milgromian potential $\Phi(\boldsymbol x)$. This time, the boundary condition 
\begin{align}
	\label{eq:bound_mond}
	\Phi(r) = \left(G M_\text{b} a_0\right)^{1/2} \ln (r)
\end{align}
is applied (Eq. 20 in \cite{FamMcgaugh}), with $r$ again being the distance to the center of baryonic mass.
This boundary condition holds true if $|\nabla\phi(r)|/a_0 \ll 1$.
\end{enumerate}

This realization of Milgromian dynamics is very efficient from a code-developing point of view, because it allows us to make use of already existing classical Poisson solvers and thus of existing grid-based codes.\footnote{The only limitation is that this scheme can not be implemented into codes which require to linearly add accelerations. This applies to a number of available codes using nested grids (e.g. the {\sc superbox} code), and also tree-codes.}
Nevertheless, a larger number of changes -- apart from adding the PDM density routine -- had to be made for technical reasons. The code will be made publicly available as a patch of \RAMSES{} in the future.

\subsection{Phantom dark matter density: the diskrete scheme}
\label{sect:PDM}
\begin{figure}
\centering
\includegraphics[width=8.5cm,bb=-20 0 480 474]{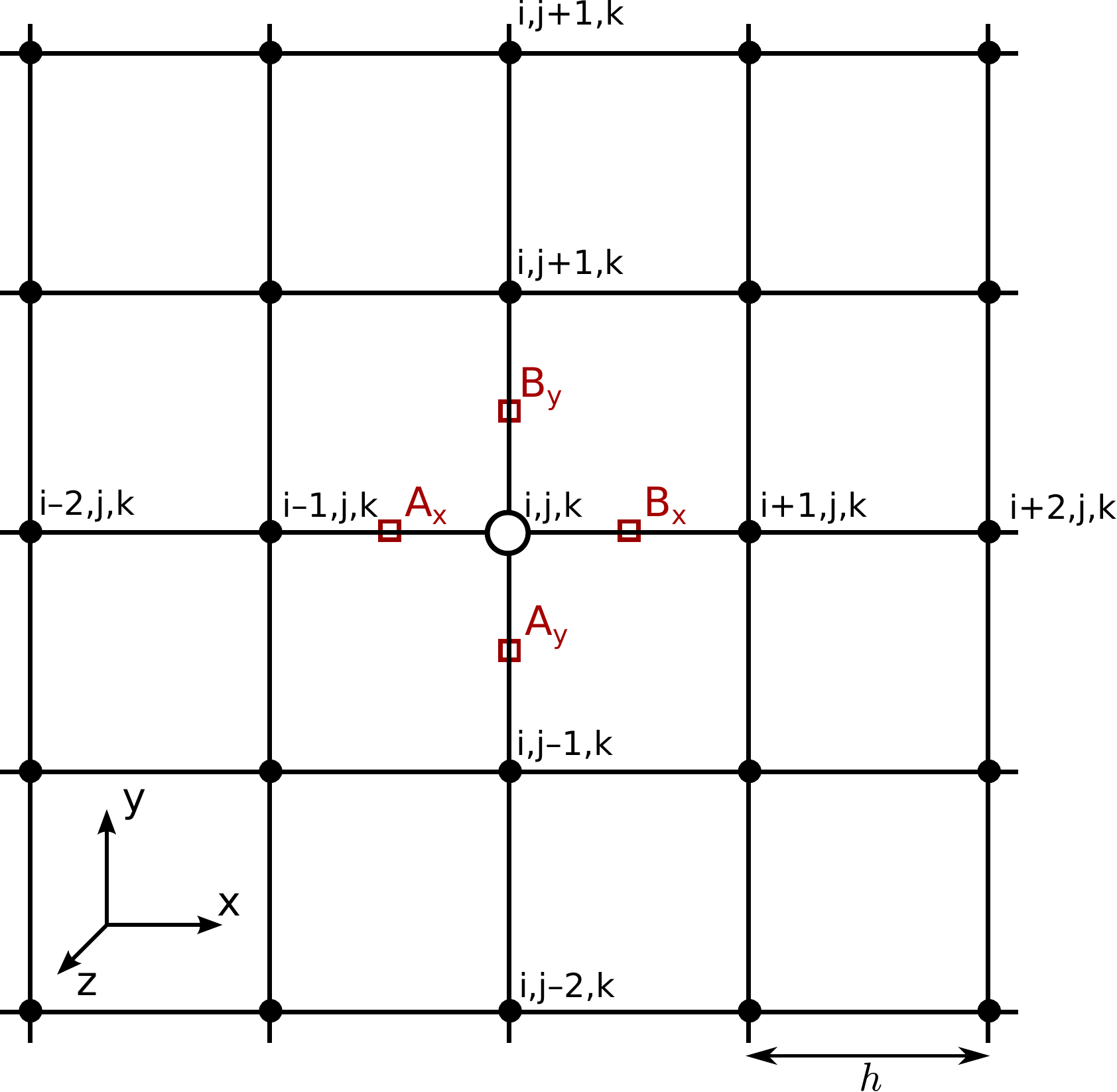}
\caption{
	Illustration of the diskretisation scheme in the x-y plane, referring to Sect.~\ref{sect:PDM}. The grid cells are of equal size with a constant cell with $h$. The values of $\nu(y)$ are evaluated at the points A and B.
}
\label{fig:grid}
\end{figure}

The PDM density (Eq.~\ref{eq:rho_phantom}) is approximated at the center of the grid cell $(i,j,k)$ at the position $\boldsymbol x_{i,j,k}$ by
\begin{align}
\rho_\mathrm{ph}^{i,j,k} = \frac{1}{4\pi G h}
		\bigg[ \ \ \  &\nu_\mathrm{B_x} \left(\nabla \phi\right)_\mathrm{B_x,x}
		 - \nu_\mathrm{A_x} \left(\nabla \phi\right)_\mathrm{A_x,x}
		 \\
		+\ & \nu_\mathrm{B_y} \left(\nabla \phi\right)_\mathrm{B_y,y}
		 - \nu_\mathrm{A_y} \left(\nabla \phi\right)_\mathrm{A_y,y}
		 \nonumber\\
		+\ &\nu_\mathrm{B_z} \left(\nabla \phi\right)_\mathrm{B_z,z}
		 - \nu_\mathrm{A_z} \left(\nabla \phi\right)_\mathrm{A_z,z} \ \ \ \bigg]
		 \nonumber \,.
\end{align}
See also \cite{PRG}.
Here, $\phi$ is the classical Newtonian potential computed in the first step, $h$ is the cell width (see Fig.~\ref{fig:grid}). 
$\nu_\mathrm{A_x}$ is the value of $\nu(|\nabla \phi|/a_0)$ at the point $\mathrm{A}_\mathrm{x}$. 
$\left(\nabla \phi\right)_\mathrm{A_x,x}$ is the x-component of $\nabla \phi$ at the point $\mathrm{A}_\mathrm{x}$ and is approximated by the following finite difference approximation: 
\footnote{We abbreviate $\phi^{i+\Delta i,j+\Delta j,k+\Delta k}$ by $\phi^{\Delta i,\Delta j,\Delta k}$.}
\begin{align}
\left(\nabla \phi\right)_\mathrm{A_x,x} &= 
	\frac{\pp{-2,0,0} -27 \pp{-1,0,0} +27 \pp{0,0,0} - \pp{1,0,0}}{24 h} 
\nonumber\\
\left(\nabla \phi\right)_\mathrm{A_x,y} &= 
	0.5 \left[ \ff{y}{-1,0,0} + \ff{y}{0,0,0} \right]
\\
\left(\nabla \phi\right)_\mathrm{A_x,z} &= 
	0.5 \left[ \ff{z}{-1,0,0} + \ff{z}{0,0,0} \right]
\nonumber
\end{align}
with
\begin{align}
\ff{y}{i,j,k} &= \frac{\pp{0,-2,0} -8 \pp{0,-1,0} +8 \pp{0,1,0} - \pp{0,2,0}}{12h} \\
\ff{z}{i,j,k} &= \frac{\pp{0,0,-2} -8 \pp{0,0,-1} +8 \pp{0,0,1} - \pp{0,0,2}}{12h} \nonumber \,.
\end{align}
This scheme applies to the other points analogously.
At the fine levels' grid boundaries, the Newtonian potential $\phi$ is interpolated from the next coarse grid level.

\section{Testing}\label{sect:testing}
The \RAMSES{} code has already been tested extensively (see \cite{Teyssier2002,GuilletTeyssier2011}).
Here, we present tests of the \POR{} code to show that the QUMOND extension works correctly and no bugs have been implanted.
We start with checking the distribution of PDM to make sure that the QUMOND routine is well implemented. 
If this test is passed, we proceed to dynamical models that are known to be dynamically stable and check the stability of these in the \POR{} code. We finally compare the results to those obtained by the \NMODY{} code \cite{nmody} using the same initial conditions.

If not stated otherwise, we apply the following $\nu$-function:\footnote{This particular $\nu$-function reproduces well the observed gravitational field in galaxies, nevertheless it fails in the Solar System \cite{Hees14}.}
\begin{equation}
	\nu(y) = \frac{1}{2} \left( 1+\frac{4}{y} \right)^{1/2} - 1 \,.
\end{equation}
The code is however not limited to this particular $\nu$-function, the implementation of this formula can be changed readily in \POR{}.

\subsection{Static tests}
The heart and soul of the \POR{} code (and the major change to the original \RAMSES{} code) is the routine that computes the PDM density from the Newtonian potential to obtain the Milgromian potential. If we assume the rest of the code to work correctly, as it has done before being customized, it should be sufficient to ensure that the PDM density and subsequently the resulting Milgromian acceleration is computed correctly at every time step.
To test this critical routine, we ran the code with models having known analytical solutions. In this paper, we present the results for a single point mass and a Plummer model.

\subsubsection{PDM density and acceleration}
\begin{figure*}
\centering
\begin{minipage}[b]{9cm} 
\includegraphics[width=8.5cm,bb=26 26 551 766]{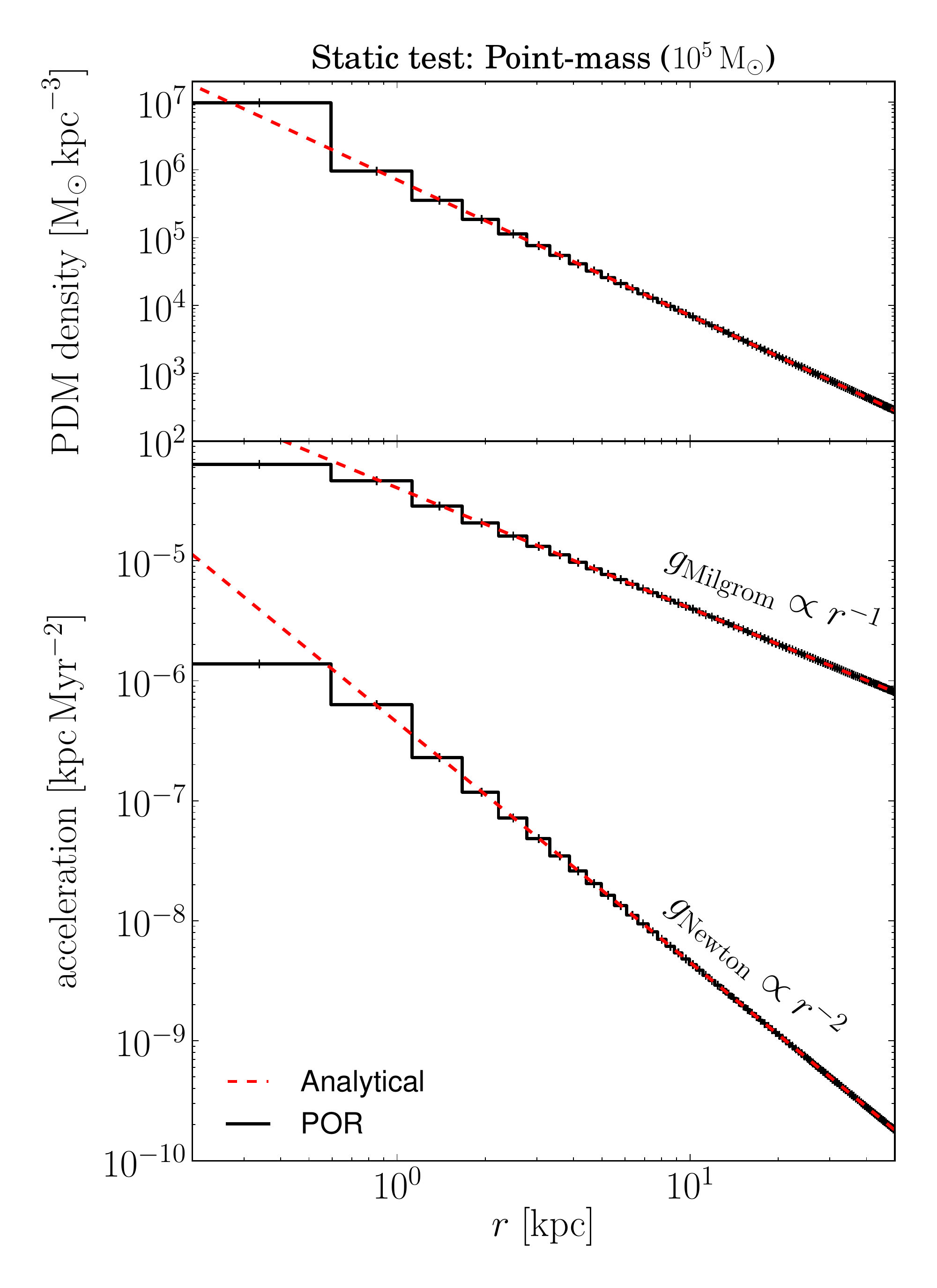}
\end{minipage}
\hfill
\begin{minipage}[b]{9cm} 
\includegraphics[width=8.5cm,bb=26 26 551 766]{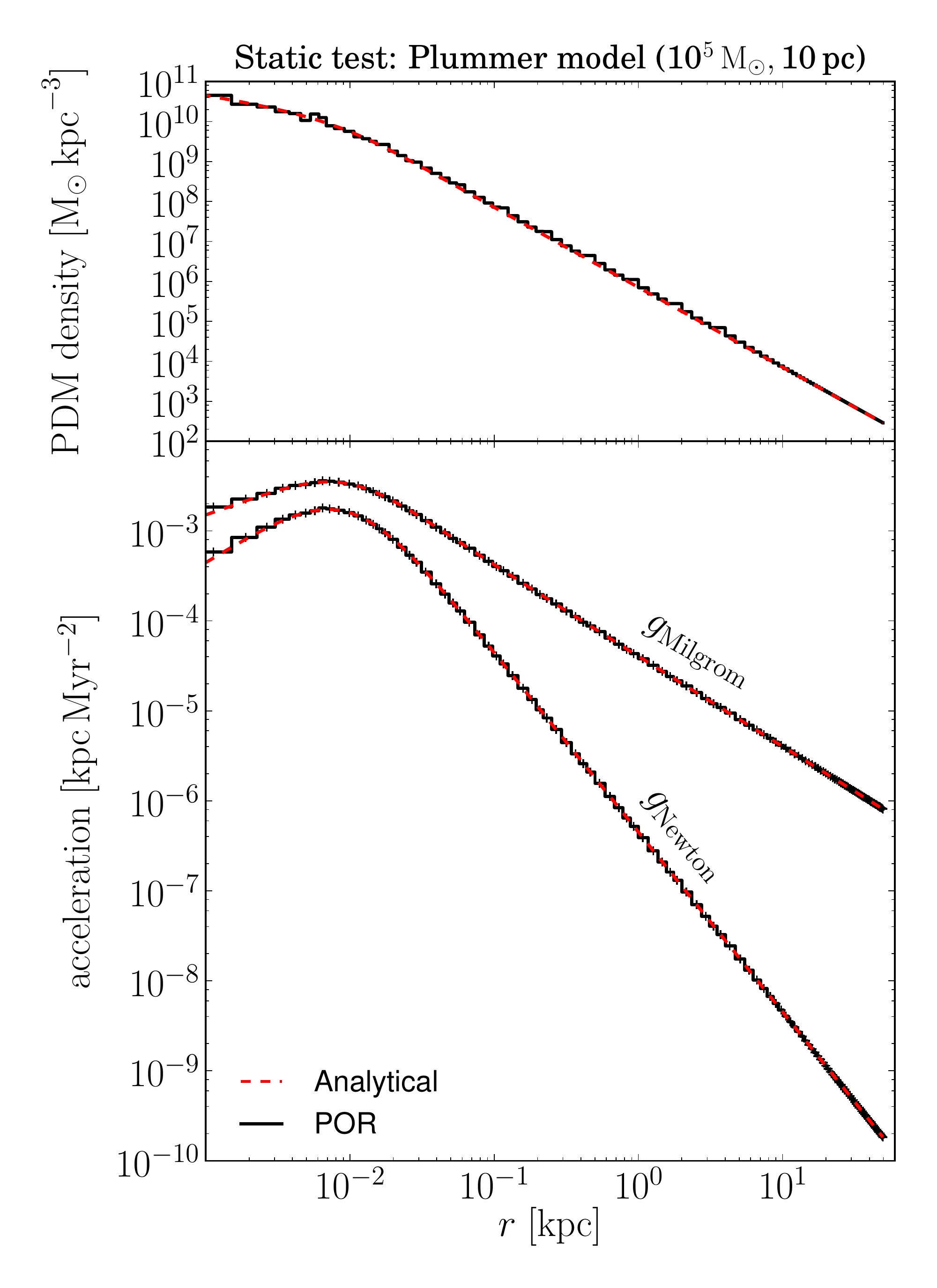}
\end{minipage}
\caption{
	The figure shows the PDM density distribution (upper panels) and the Newtonian and Milgromian accelerations (lower panels) computed by the \POR{} code for (a) a point mass (left-hand side) and (b) a Plummer model (right-hand side). The steps (black solid lines) represent the values of individual cells along the diagonal ($x=y=z>0$). The ticks on the steps mark the cell centers. The red dashed lines are derived analytically. Both models are in the deep Milgromian regime, i.e. $g_\mathrm{N} \ll a_0$.
	The grid resolution of case (a) is fixed to $2^8$ cells at a box length of 100\,kpc, i.e. 0.39\,kpc. The grid resolution of model (b) is variable and ranges from 0.39\,kpc (coarsest grid level, i.e. outer regions) to 0.76\,pc (finest grid level, i.e. box center), the box size is again $(100\unit{kpc})^3$.
}
\label{fig:test-PDM-and-acceleration}
\end{figure*}
We set up two spherical models: 
a) a point-mass with $10^5\,\Msun$ which is located at the box center. 
And b) a Plummer model with a total mass of $10^5\,\Msun$ and a Plummer radius of 10\,pc. The Plummer model is populated with $10^5$ particles.
We execute the code using the following parameters:  \\
- simulation box size: (100\,kpc)$^3$, \\
- minimum resolution: $100\unit{kpc}/2^8 = 0.39 \unit{kpc}$, \\
- maximum resolution: $100\unit{kpc}/2^{17} =  0.76  \unit{pc}$.\\
The resolution of the point-mass model is fixed to 0.39\,kpc. The grid of the Plummer model is adaptively refined.
We plot the resulting PDM density distribution as a function of radius along the diagonal ($x=y=z > 0$) as well as the resulting effective acceleration $g$ in Fig.~\ref{fig:test-PDM-and-acceleration}.
Both, $\rho_\mathrm{ph}$ and $g$, agree well with the analytical solution. 
The PDM density shows little scatter, particularly at the level boundaries (top right panel in Fig.~\ref{fig:test-PDM-and-acceleration}). This is because the second order of $\phi$ enters in Eq.~\ref{eq:rho_phantom}. This scatter is however averaged out in the final acceleration, $g$.
In the case of the point-mass model which has a fixed cell width, one can see that the acceleration $g_\text{N}$ is smoothed in the innermost cell.\footnote{The effective resolution of such a grid code is approximately twice the fine level cell width.} As a side effect, the PDM density is slightly off in this cell. In a model made of particles or gas, the smoothing is however negligible if the resolution is sufficient, which is generally the case and can well be seen in Fig.~\ref{fig:test-PDM-and-acceleration}.

For comparison, also the Newtonian acceleration $g_\mathrm{N}$ is plotted and compared to its analytical pendant to stress the quality of the resulting Milgromian acceleration $g$.

\subsubsection{Dependence on spatial resolution}
The \RAMSES{} code works with a so called one-way interface scheme \cite{Kravtsov97,Teyssier2002}. This means it starts solving the Poisson equation at the coarsest level, and uses this coarse level solution of the potential as the boundary condition at the next finer grid level, but not vice versa. I.e., it does not use the solutions of the finer levels to correct the coarse level solution (this would be referred to as a two-way interface scheme). For the \POR{} code, this means that the PDM density distribution on the coarse level alone defines this level's Milgromian potential. 
A lack of resolution of the PDM density could however result in a non-accurate acceleration on the not so well refined grid level, particularly on the coarse grid level.

However, in Fig.~\ref{fig:test-PDM-and-acceleration} one sees no such dependence on the resolution.
To quantify this finding, we compare the PDM mass of the coarse grid to the PDM mass of the completely refined grid (i.e. of the leaf cells). In order to get a meaningful result, we only consider the regions that are refined, i.e. are having sub-cells. 
We find that the difference between the PDM mass of the coarse level cells and the related leaf cells is tiny. For the presented Plummer model, the relative difference is $\approx 4\times 10^{-5}$, although the spatial resolution of the finest grid is 32 times as high as the coarse grid resolution.

\subsection{Dynamical tests}
\label{sect:dynamical_tests}
As the PDM density, $\rho_\mathrm{ph}$, and the resulting Milgromian acceleration, $g=-\nabla \Phi$, have been shown to perform well, we proceed to evolving the Plummer model dynamically. To do so, we set up initial conditions that are, in Milgromian dynamics, in dynamical equilibrium.
To find such initial conditions, we use the following distribution function (Eddington's formula \cite{EddingtonFormula}).
\begin{align}
f(E) = \frac{1}{\sqrt{8} \pi^2} \left[ 
	\int_0^E \frac{\d\Phi}{\sqrt{E-\Phi}} \frac{\d^2 \hat\rho_\mathrm{b}}{\d \Phi^2} + \frac{1}{\sqrt{E}}\left(\frac{\d\hat\rho_\mathrm{b}}{\d\Phi}\right)_{\Phi=0}
\right]
\end{align}
with $\hat\rho_\mathrm{b}(r)$ being the normalized density distribution of baryonic matter, 
$\Phi(r)$ the Milgromian potential energy, 
and $E$ the kinetical energy.
$\Phi(r)$ is computed from the baryonic matter density plus PDM density, $\rho_\mathrm{b}(r) + \rho_\mathrm{ph}(r)$, by solving Eq.~\ref{eq:poisson2} in spherical coordinates.
The spherical PDM density distribution,
 \begin{equation}
 \rho_\mathrm{ph}(r) = \frac{1}{4 \pi G} \frac{1}{r^2} \frac{\d}{\d r} \left( r^2\  \nu(|\d\phi/\d r|/a_0) \frac{\d\phi}{\d r} \right) \,,
 \end{equation}
 has been computed numerically for this symmetric model.

\subsubsection{Stability of a spherical model in dynamical equilibrium: Lagrangian radii and radial velocity dispersion}
\label{sect:plummer-lagrangeradii-velocitydispersion}
\begin{figure}
\centering
\includegraphics[width=8.0cm,bb=15 8 430 494]{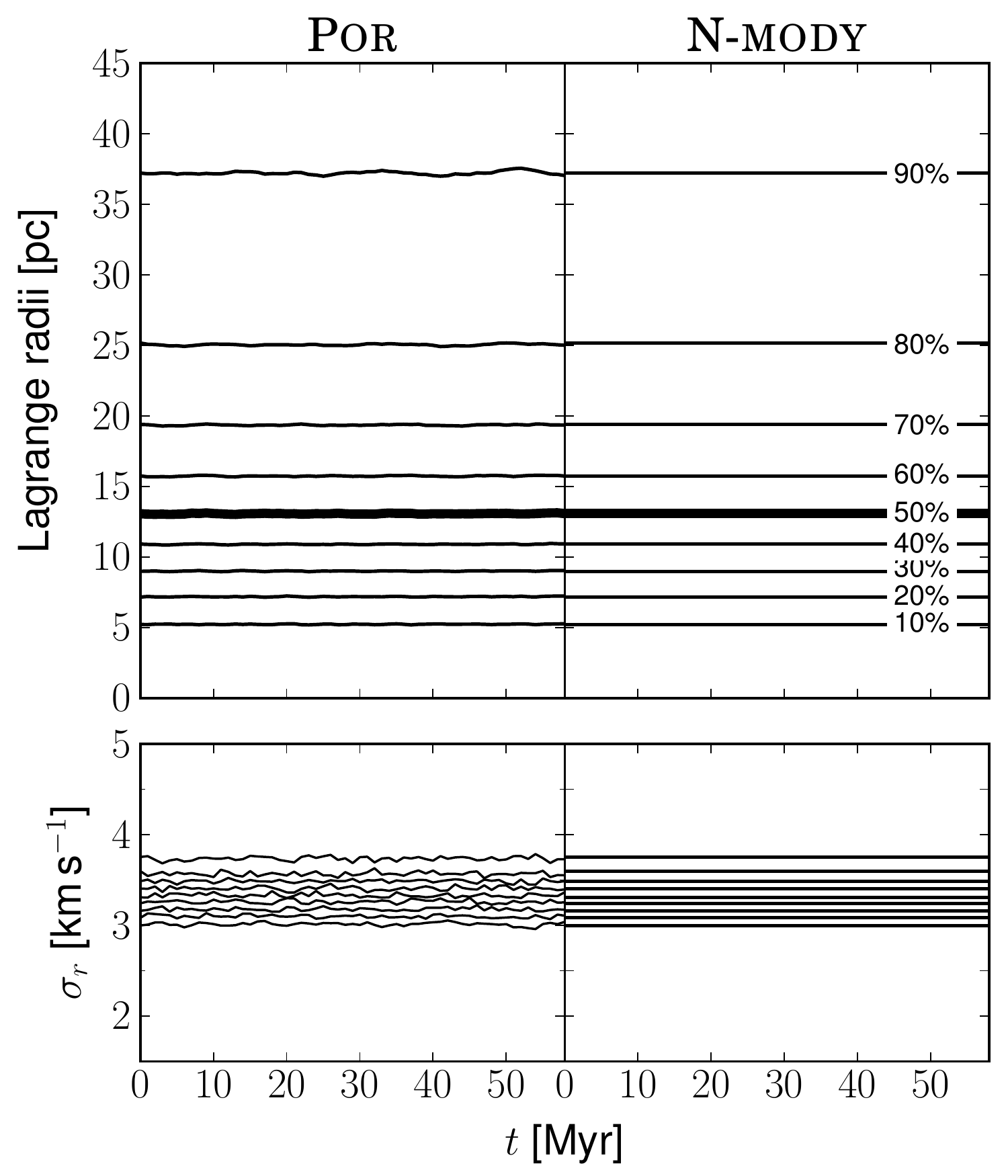}
\caption{
	The stability of a Plummer model is tested and compared to \NMODY{}.
	The Lagrangian radii (10\%, 20\%, \dots, 30\%) and radial velocity dispersion (in the radial bins defined by the Lagrangian radii, 0--10\%, 10--20\%, \dots, 80--90\%) are plotted vs. time. The Plummer model has a Plummer radius of $10\,\textrm{pc}$ and a total mass of $10^5\,\Msun$ (deep Milgromian regime). The $N$-body model is made of $10^5$ particles, and exactly the same initial conditions are used for both runs (\POR{} and \NMODY{}). See Sect.~\ref{sect:plummer-lagrangeradii-velocitydispersion} for details. The plot covers a time range of more than 100 crossing times of this system.
}
\label{fig:plummer-lagrangeradii-velocitydispersion}
\end{figure}
We evolve the Plummer model using the \POR{} code and plot the Lagrangian radii (10\%, 20\%, \dots, 90\%) vs. time in Fig.~\ref{fig:plummer-lagrangeradii-velocitydispersion} (left hand side panels).
We find that the Lagrangian radii and consequently the density profile stay constant with time.

Moreover, we compute the radial velocity dispersion within the radial bins defined by the Lagrangian radii (0--10\%, 10--20\%, \dots, 80--90\%) and plot them in the bottom left panel of Fig.~\ref{fig:plummer-lagrangeradii-velocitydispersion}. Also the radial velocity dispersions do not change with time, meaning that the test model is shown to be dynamically stable, as expected from theory.

\subsubsection{Comparison to the N-MODY code}
We execute the \NMODY{} code using exactly the same initial conditions we used in Sect.~\ref{sect:plummer-lagrangeradii-velocitydispersion}.
The simple $\nu$-function, $\nu(y) = 0.5 (1+4/y)^{1/2}$, is used in the \POR{} code and the related $\mu$-function, $\mu(x) = x/(1+x)$, in \NMODY{}.
We find that the results of both codes are in agreement (see Fig.~\ref{fig:plummer-lagrangeradii-velocitydispersion}, compare the panels on the left hand side to those on the right).
The radial velocity dispersion appears to be more stable in \NMODY{}. This is due to the spherical grid architecture, which is ideally suited for such spherical problems and allows resolving the spherical potential better than the Cartesian cell-based grid in the \POR{} or \RAMSES{} code.

\subsubsection{Conservation of energy}

\begin{figure}
\centering
\includegraphics[width=8.0cm,bb=25 9 531 423]{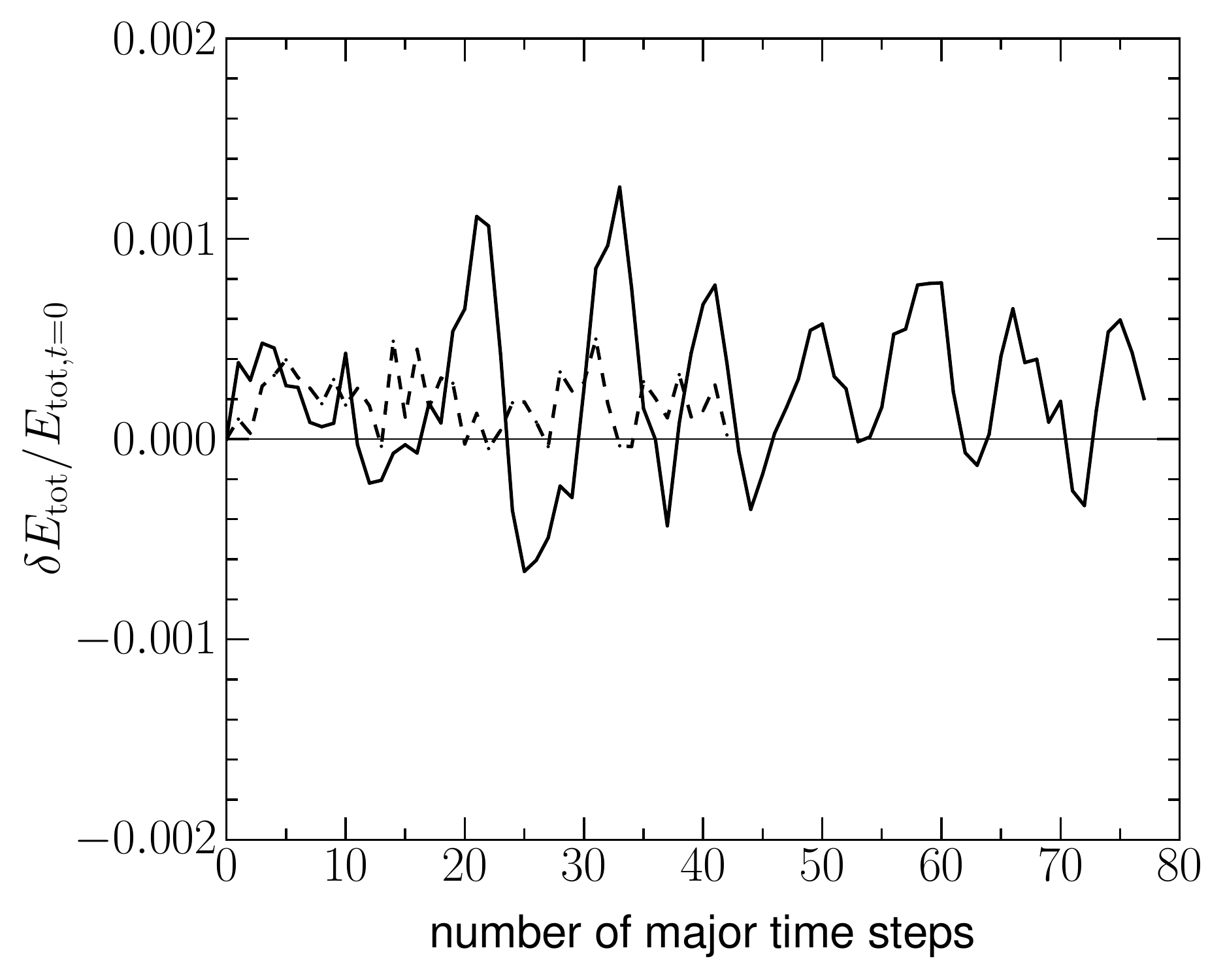}
\caption{
	The relative energy error is visualized for the Plummer model described in Sect.~\ref{sect:testing}. The solid curve shows the relative energy error from the \POR{} code. As a benchmark, we plot this error as the dashed curve also for the \RAMSES{} code (using a Plummer model with same size/mass parameters, but with initial velocities yielding virial equilibrium in Newtonian dynamics).
	One course time step is $\approx 1.24\unit{Myr}$ in the Milgromian model and $\approx 2.34\unit{Myr}$ in the Newtonian one.
}
\label{fig:econs}
\end{figure}

Because the QUMOND Poisson equation (Eq.~\ref{eq:poisson}) is derived from an action with a modified Lagrangian (see \cite{QUMOND}, and Sect. 6.1.3 in \cite{FamMcgaugh}), the related equations of motion obey conservation of energy and angular momentum. 

At each time step, we compute the total kinetic energy,
\begin{equation}
	T = \sum_{i=1}^{N_\textrm{part}} m_i v_i^2,
\end{equation}
of the $N_\textrm{part}$ particles having the masses $m_i$ and the velocities $v_i$. 
The virial, $W$, (with units of potential energy) is computed on the grid (leaf) cells,
\begin{equation}
	W = -\int\limits_{\mathbb{R}^3} \rho(\boldsymbol x) \left( \boldsymbol x \cdot \nabla \Phi(\boldsymbol x) \right) d^3x
		= \sum_{i=1}^{N_\textrm{cells}} \rho_i h_i^3 \left( \boldsymbol x_i \cdot \boldsymbol g_i \right)
\end{equation}
(see \cite{BT}),
where $\rho_i$ is the baryonic mass density distribution and $\Phi_i$ the Milgromian gravitational potential at the center of the $i$-th cell (with the cell width $h_i$). $\boldsymbol g_i = -\nabla\Phi_i$ is the acceleration.\footnote{The integration over particles instead of grid cells, $W = -\sum_{i=1}^{N_\textrm{part}} m_i \left( \boldsymbol x_i \cdot \boldsymbol g_i \right)$, with $m_i$ being the mass of the $i$-th particle, which is located at position $\boldsymbol x_i$ and feels the acceleration $\boldsymbol g_i$, yields the same result, but is computationally more expensive.}

We print out the change of total energy, $\delta(T+W)$, with time, the so called numerical `energy error', at each time step.
This energy error always plays a role in numerical experiments, because the complexity of reality forces us to make simplifications to keep the numerical effort at a feasible level, inevitably resulting in such numerical errors.
We find that the energy is well conserved within the common tolerances and on approximately the same level (although very slightly higher) as the original or Newtonian \RAMSES{} code.
To back up the latter statement, we set up a Plummer model similar to that described here in Sect.~\ref{sect:testing}, but under the assumption of Newtonian dynamics to be valid. In this Newtonian model, the virial is approximately half of that of the Milgromian model. We evolve this model in the Newtonian \RAMSES{} code and plot the resulting energy error of both codes in Fig.~\ref{fig:econs}.
The average amplitude of the energy errors of both codes are similar. It is however noticeable that the variation of the error is slightly larger in the Milgromian code, which is due to the fact that the numerical error induced by the computation of the PDM density adds to the energy error.

\subsubsection{General note on the overall code efficiency: RAMSES vs. POR}
The title of this subsection could also have been ``simple Poisson solver + DM particles vs. more sophisticated Poisson solver without the need of DM particles''.
On the one hand, this particular implementation of Milgromian dynamics requires solving the Poisson equation two times instead of only once per time step, making the force computation less efficient compared to the classical Newtonian Poisson solver of \RAMSES{}, assuming that both codes run with the same number of particles.
On the other hand, Milgromian dynamics does not require any DM particles. If we take into account that a large fraction of particles used in galactic dynamics computations are DM particles, this advantage compensates (and even overcompensates) the handicap that comes with the more expensive QUMOND solver.
Computations with the same number of baryonic particles are therefore generally faster in the DM-free framework.


\section{Application}
\label{sect:applications}
\begin{figure}
\centering
\includegraphics[width=8.2cm,bb=30 23 530 409]{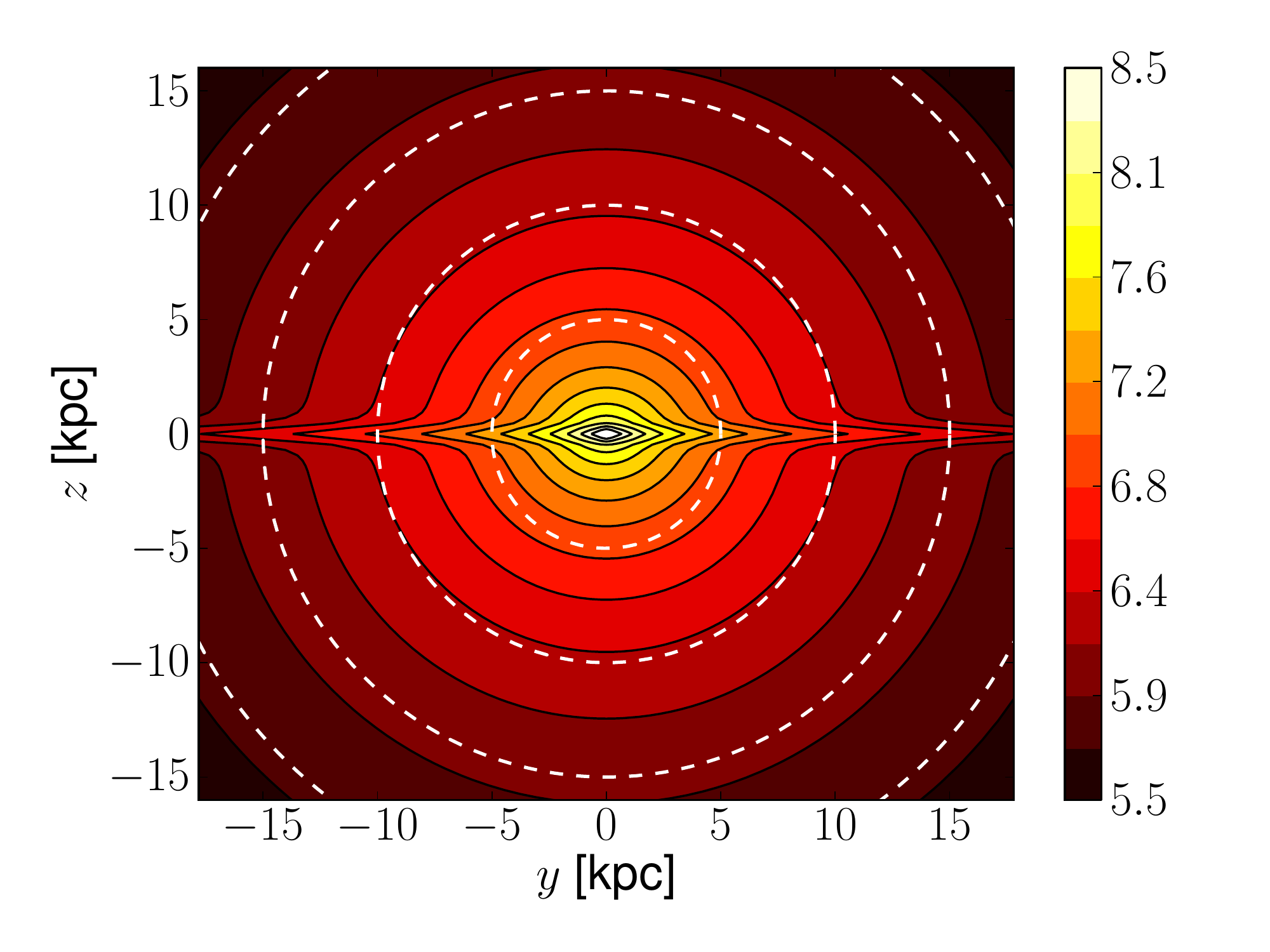}
\caption{
	Phantom dark matter (PDM) density distribution of the exponential disk model described in Sect.~\ref{sect:disk} (logarithmic scale), $\log_{10}\left( \rho_\mathrm{ph} / \left(\mathrm{M}_\odot \,\mathrm{kpc}^{-3}\right) \right)$. 
	The PDM distribution of a disk typically features a close-to-spherical (slightly oblate) halo component which has an isothermal density profile at large radii, $\rho_\mathrm{ph}(r) \propto r^{-2}$, and a dark disk component which is aligned with its baryonic counterpart, but has a different scale length.
	The phantom dark disk is one of the outstanding differences to the standard DM halo model.
}
\label{fig:diskpdmhalo}
\end{figure}

Beside the testing of the \POR{} code, we demonstrate its functionality on the basis of basic dynamical applications. 
Besides the spherical equilibrium models in Sect.~\ref{sect:dynamical_tests}, these applications are (a) rotating disk galaxies and (b) disk galaxies with an initial bulge.
We developed the software tools necessary to set up these components in Milgromian dynamics, which is generally not trivial. 
Having these basic tools at hand, we can construct initial conditions for more sophisticated models, e.g. galaxy-galaxy interactions, dwarf spheroidal galaxies orbiting a host galaxy, etc. 


Disk galaxies, as rotation-supported objects, are known to be well described by Milgromian dynamics \cite{FamMcgaugh}.
These objects obey the baryonic Tully-Fisher relation,
\begin{equation}\label{eq:btfr}
	V_\text{flat}^4 = G a_0 M_b
\end{equation}
with $a_0$ being Milgrom's constant. This relation follows directly from Eq.~\ref{eq:milgromsformula}, and holds true over more than five orders of magnitude in mass \cite{Verheijen2001,StacyPRL,McGaugh2012}.
The observation of this relation is however in contradiction with the DM-hypothesis, because it in fact states that the flat rotation velocity, $V_\mathrm{flat}$, depends only on the total baryonic mass, although the mass of the dark matter does, in the standard model, constitute the largest fraction of the total mass ($\approx$ 80--90\%). What is more, $V_\mathrm{flat}$ does not depend on the surface density. Given the available large statistical sample of galaxies with rotation curves (e.g. more than 100 objects analyzed by \cite{StacyPRL}), this conflict with DM-based theories can be solved only with a large and thus unlikely amount of fine-tuning.

Also the mass discrepancy-acceleration (MDA) relation predicted by Milgromian dynamics, relating the observed rotation velocity $V_\text{obs}$ to the theoretical rotation velocity $V_\mathrm{b}$ as deduced from the distribution of baryonic matter under the assumption of Newtonian gravity,
\begin{equation}\label{eq:mda}
	V_\text{obs}^2 / V_\text{b}^2 = M_\text{dyn} / M_\text{b} = \nu(|g_\mathrm{N}|/a_0) \,,
\end{equation}
is significantly visible in the data of rotation-supported galactic systems \cite{Sanders1990,Scarpa2006,McGaugh2012}.
In this relation, $M_\mathrm{b}(r)$ is the baryonic mass enclosed in the galactocentric radius $r$, and 
$M_\mathrm{dyn}(r) = V_\mathrm{obs}^2(r) r /G$
is the related enclosed `dynamical' mass inferred from the observed rotation velocity at radius $r$ under the assumption of Newtonian dynamics to be valid.
In the standard model, this dynamical mass would be interpreted as sum of the masses of baryonic and dark matter.
Particularly the fact that the MDA relation includes also individual wiggles in the rotation curves is interesting, because such wiggles would be washed out by a existing CDM halo.
The existence of either of these two relations, involving the {\it same} acceleration constant $a_0$, is not understood in the standard model, and is without reasonable argumentation claimed to emerge from not yet explored galactic feedback processes \cite{Foreman2012,McGaugh2011Foreman}. On the other hand, these relations are natural consequences of the Milgromian framework. The observational data is perfectly in agreement with these relations, thus motivating to explore the Milgromian framework and its consequences in more detail.

To date, most predictions are based on static models simply using Eq.~\ref{eq:milgromsformula}.
Only few studies investigated the dynamical evolution of galaxies, and so the dynamical consequences of this DM-free framework are widely unexplored. 
The stability of stellar disks has been studied by \cite{BradaMilgrom1999} and more recently by \cite{Tiret2007}.
As expected, the recent study shows that exponential disk galaxies qualitatively behave dynamically differently in the Newtonian+DM vs. Milgromian frameworks. While in the standard model disks are stabilized by pressure-support haloes made of dissipationless DM particles, this is not the case in the Milgromian framework in which self-gravity plays a much larger role. It has been shown that disks are, in Milgromian dynamics, more sensitive to instabilities resulting in the formation of galactic bars. While galactic bars grow slowly in a DM-stablized halo, they form quickly in a DM-free Milgromian universe. 
Also dynamical friction behaves very differently. When the bar has formed in the DM-based model, it is subjected to dynamical friction with the DM particles, thus exchanging angular momentum and slowing down its pattern speed. This is not the case for the DM-free Milgromian dynamics, where the pattern speed stays constant for many Gyr. 

Another difference between both models is that the Milgromian disk potential has, contrary to close-to-spherical particle DM haloes, a phantom dark disk component which is aligned with its baryonic counter-part (see Fig.~\ref{fig:diskpdmhalo}), providing e.g. a stronger azimuthal force than the baryonic component alone. 
The scale lengths and heights of the phantom dark disk and the baryonic disk are however generally different.

Here, we create $N$-body models of disk galaxies, with and without a bulge, and compare them to those investigated in \cite{Tiret2007} to demonstrate the performance of this code.
While our computations are based on the QUMOND formulation, \cite{Tiret2007} base their work on a different Milgromian dynamics theory, AQUAL \cite{BM84}. Although both theories are similar, small quantitative differences are expected.

\subsection{Rotating stellar disk}
\label{sect:disk}

\begin{figure}
\centering
\includegraphics[width=8.5cm,bb=25 30 565 415]{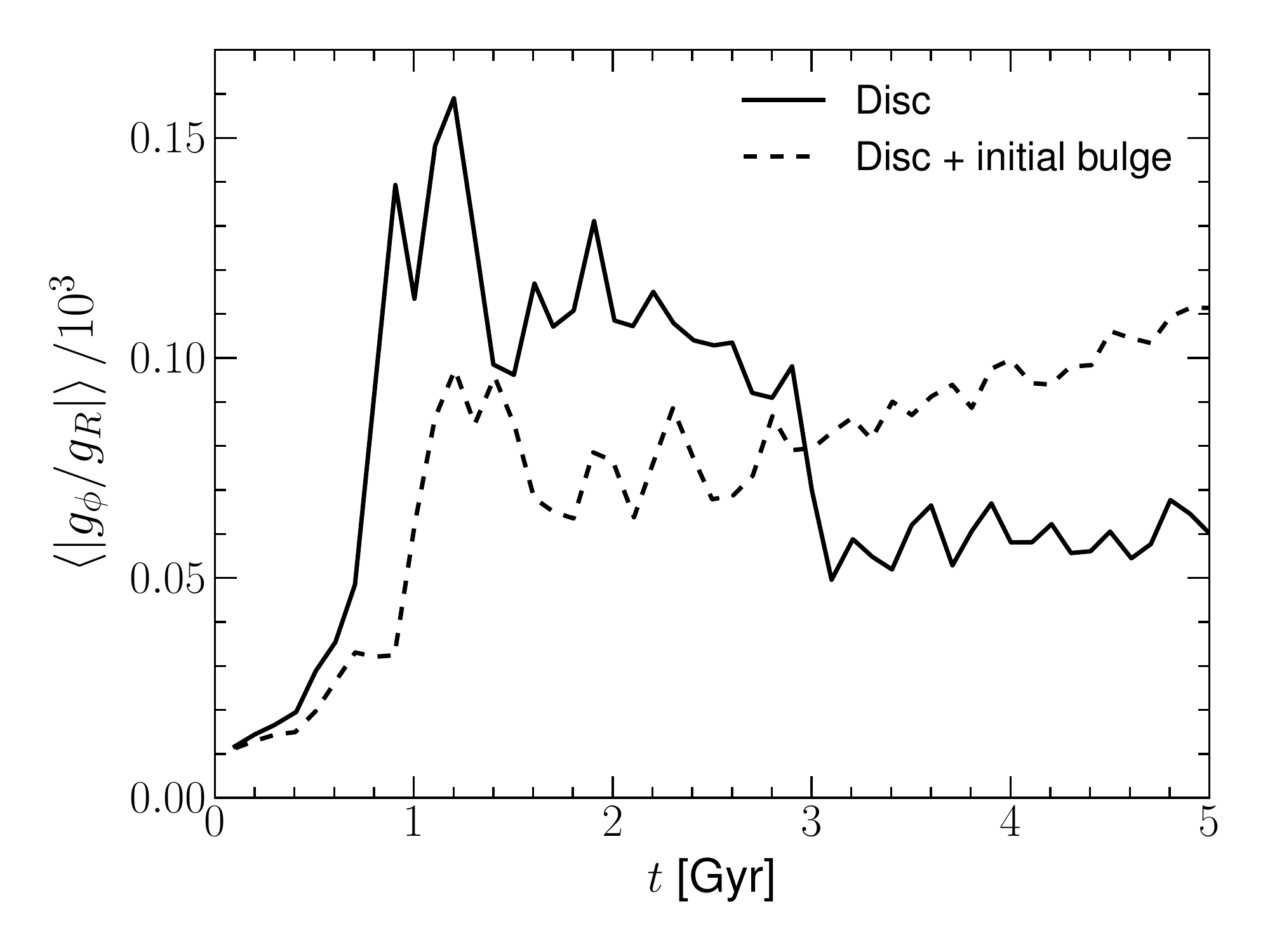}
\caption{
	Evolution of the galactic bar strength or asymmetry of the galactic potential, $\left< \left| g_\phi / g_\mathrm{R} \right| \right>$ (see Eq.~\ref{eq:barstrength}). In the case of the disk-only model, the bar grows within 1\,Gyr, remains until $t\approx 3\unit{Gyr}$, and finally partly dissolves. What remains is a typical peanut-shaped galaxy.
	In the case of the disk model with initial bulge, the asymmetry of the galactic potential grows more slowly and continuously. 
	The most prominent peak, which occurs between $1\unit{Myr}$ and $1.5\unit{Myr}$ is due to the formation of spiral arms.
}
\label{fig:barstrength}
\end{figure}

\begin{figure*}
\centering
\includegraphics[width=17.5cm,bb=15 14 705 382]{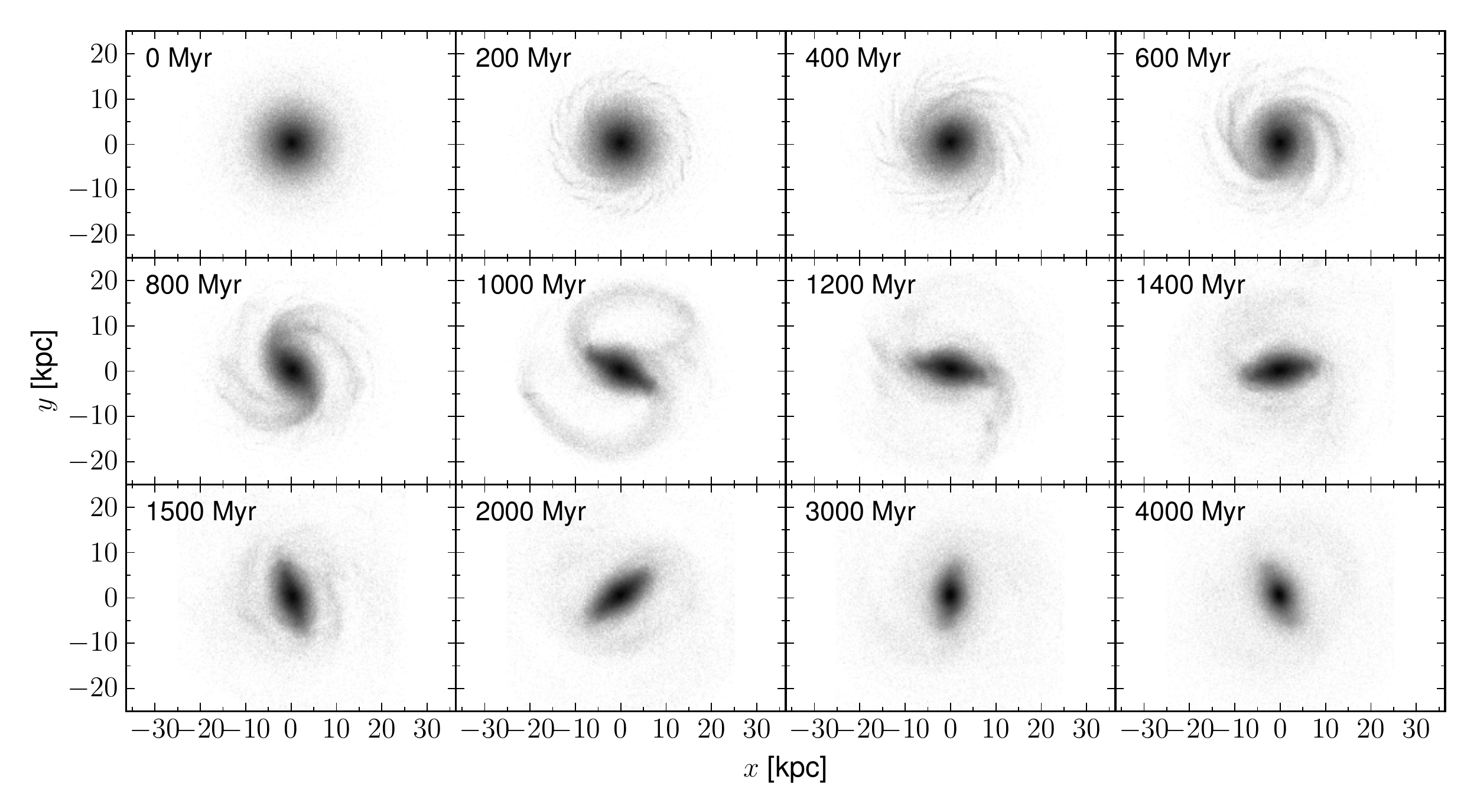}
\caption{
	Evolution of an exponential disk with time in the new Milgromian $N$-body code, \POR{}.
	The disk contains only stellar particles and is not embedded in a dark matter halo.
	It forms substructure within a few 100\,Myr due to self-gravity. From these instabilities, a galactic bar grows within $\approx 1\unit{Gyr}$. This bar rotates with a constant angular frequency for several billion years.
}
\label{fig:diskevolution}
\end{figure*}

\begin{figure*}
\centering
\includegraphics[width=17.5cm,bb=15 14 705 382]{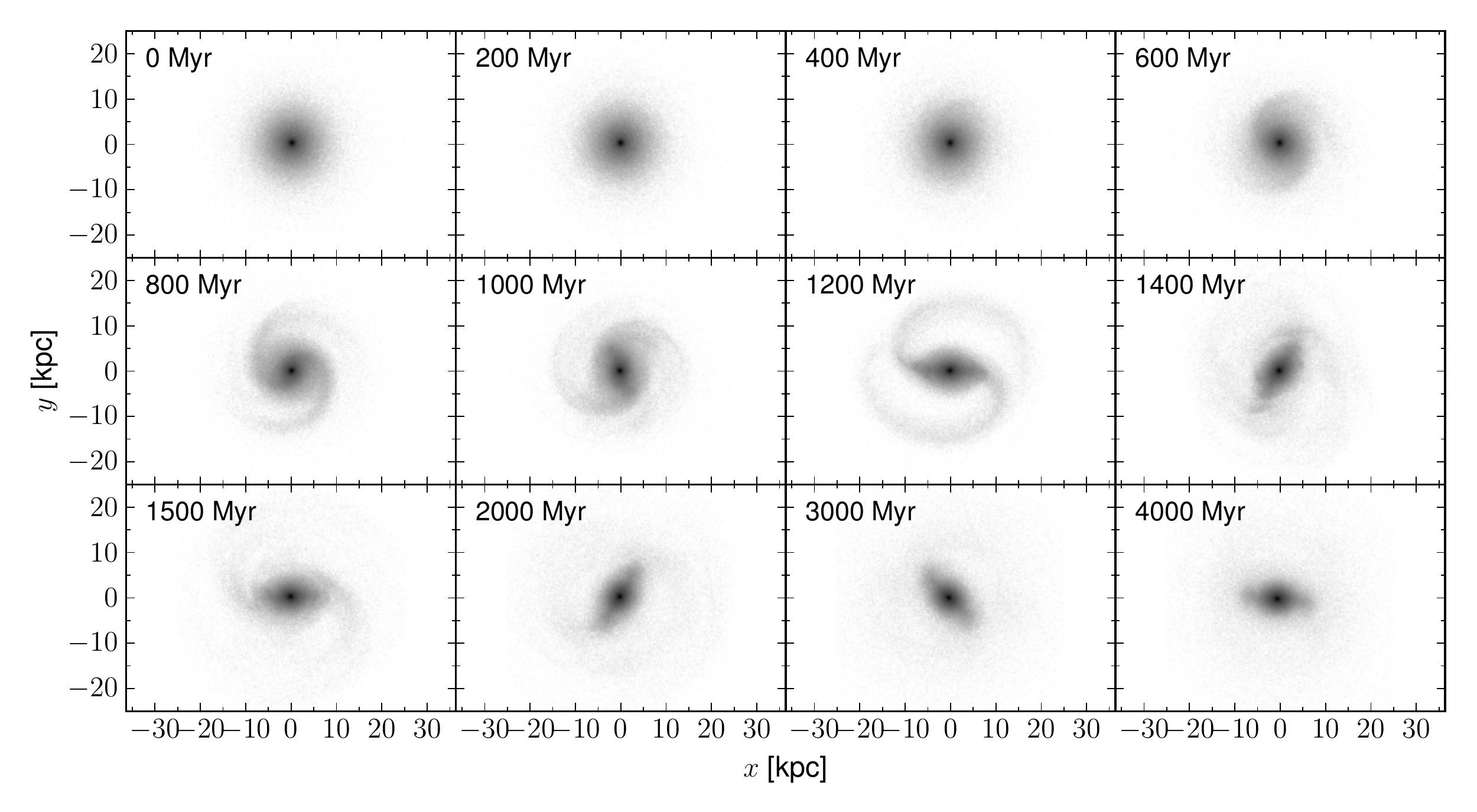}
\caption{
	Evolution of an exponential disk + bulge model with time in the new Milgromian $N$-body code, \POR{}.
	This model has a bulge that makes up $\approx 10\%$ of the total mass.
	The bulge has a stabilizing effect on the disk and delays the formation of substructure due to self-gravitation. Particularly in the beginning the formation of small-scale substructure is suppressed. 
	The full-length video of this computation is available as an online attachment to this paper and can also be viewed at youtu.be/GhR5w7qDTX8.
}
\label{fig:diskbulgeevolution}
\end{figure*}

We set up a rotating exponential disk having the density distribution
\begin{equation}\label{eq:expdisk}
	\rho_\mathrm{b,disk}(R,z) = \rho_0 \exp(-R/R_\mathrm{d}) \, \mathrm{sech}^2(z/z_0) \,,
\end{equation}
with $R_\mathrm{d}$ being the radial scale length and $z_0$ the azimuthal one.
The model is initially in virial equilibrium. 

To find matching initial conditions, we adapt the already available software by W.~Dehnen, which is available with the \NEMO{} \cite{NEMO} software package and which is detailed in \cite{WD99,McMillanDehnen2007}. 
In the first step, we solve Eq.~\ref{eq:poisson} with the density distribution $\rho$ given by Eq.~\ref{eq:expdisk} using the \POR{} code. We store the solution of the Milgromian potential and acceleration in a temporary file and provide it in the next step to the galaxy setup routine as external potential, and thereby replace the potential of the DM halo by the Milgromian equivalent.

We choose the radial velocity dispersion using Toomre's stability criterion \cite{Toomre1964},
\begin{equation}\label{eq:sigr1}
	\sigma_\textrm{r}(R) = Q\,\Sigma_\mathrm{crit}(R) = Q \frac{3.36 \, G \, \Sigma_\mathrm{b}(R)}{\kappa(R)} \,.
\end{equation}
$Q$ is the Toomre parameter, $G$ the gravitational constant, $\Sigma_\mathrm{b}(R)$ the surface density at galactic radius $R$, and $\kappa(R)$ the epicyclic frequency at the same radius.
$Q$ is supposed to be constant everywhere.
$\Sigma_\mathrm{crit}$ is the so called critical surface density.
The tangential velocity dispersion is set to
\begin{equation}
	\sigma_\theta(R) = \frac{\kappa(R)}{2 \Omega(R)} \sigma_r(R) 
\end{equation}
(e.g., \cite{Hernquist1993,Tiret2007}), 
where $\Omega(R)$ is the angular frequency.
The azimuthal velocity dispersion is defined by 
\begin{equation}
	\sigma_\textrm{z}(R) = \left(R_\mathrm{d} \pi G \Sigma(R) \right)^{1/2} 
\end{equation}
(e.g., \cite{Hernquist1993,Tiret2007}).
We initialize this model with $10^6$ particles, total baryonic disk mass $M_\mathrm{d} = 5.4\ttt{10}\,\Msun$, $R_\mathrm{d} = 2.3\unit{kpc}$, $z_0 = 0.1 R_\mathrm{d}$, and $Q=2$.\footnote{This model has been inspired by the Milky Way, it is however not intended to fit the original exactly.}
The disk is placed at the center of a box of the size $(1000\unit{kpc})^3$.
The minimum grid resolution is $1000\unit{kpc} / 2^7 = 7.8 \unit{kpc}$, the maximum resolution is limited by the number of particles. In this case, this is $1000\unit{kpc} / 2^{15} = 0.031 \unit{kpc}$.

We let the model evolve for a few billion years and visualize the results in Fig.~\ref{fig:diskevolution}.
The disk starts forming substructure quickly in the outer regions where the velocity dispersion is relatively small compared to the inner regions. After $\approx 500\unit{Myr}$, the disk forms larger substructures in the form of spiral arms. And within $1\unit{Gyr}$, it happens to form a rotating galactic bulge, which rotates with a constant angular frequency. After two more billion years, the bar weakens and a central spherical component forms, giving the galaxy the typical peanut-like shape. 

To compare to the results of \cite{Tiret2007}, we compute at each time step the average absolute value of the ratio of azimuthal acceleration, $g_\phi$, to radial acceleration, $g_\textrm{R}$,
\begin{equation}\label{eq:barstrength}
\left< \left| g_\phi / g_\mathrm{R} \right| \right> = \frac{1}{N_\mathrm{cells}} \sum\limits_{i=1}^{N_\mathrm{cells}} \left|\frac{g_\phi}{g_\mathrm{R}}\right|_{i\textrm{-th cell}} \,,
\end{equation}
assuming that all cells are of equal size.\footnote{In Eq.~\ref{eq:barstrength}, we restrict to cells that have a distance to the galactic center of $1.5\unit{kpc} < R < 25\unit{kpc}$ and which are located close to the galactic plane, $|z| \leq 0.5\unit{kpc}$. These parameters appeared to give a useful description of the bar strength.} Note that this definition is not exactly the same as the one applied by \cite{Tiret2007}, but it allows a meaningful comparison, keeping in mind that also the models are not exactly the same. The quantity defined by Eq.~\ref{eq:barstrength} is zero in spherical or cylindrical symmetry and increases when the bar forms. It is a useful indicator of the strength of the galactic bar, or more generally the strength of non-ax\-isymmetries.
We plot this quantity vs. time in Fig.~\ref{fig:barstrength}. In this plot, we see what we already noticed visually: the asymmetry grows quickly within $\approx 1\unit{Gyr}$, stays until $t\approx 3\unit{Gyr}$, and then partly dissolves again. This is also what we see in figure~21 of \cite{Tiret2007}.

\subsection{Adding a bulge to the rotating disk}
\label{sect:bulge}
Setting up a spheroidal bulge in a non-spherical potential is not trivial.
One method to address this challenge is described in \cite{McMillanDehnen2007}. 
The bulge is initially set up in the spherical monopole expansion of the axissymmetric disk potential, and is then evolved in a $N$-body simulation, whereby the external spherical monopole potential is adiabatically transformed into the final non-spherical disk potential, to let the bulge settle in the actual potential of the axisymmetric galactic disk.

In the Milgromian framework, this is even more tricky, because the superposition of gravitational potentials is not applicable. To work around this limitation, we make the assumption that the Newtonian acceleration is, at the center of the galactic model, much larger than $a_0$, so that the dynamics breaks down to simple Newtonian dynamics locally. We thus set up the bulge as described in \cite{McMillanDehnen2007} under the assumption of Newtonian dynamics to be valid.\footnote{If in Eq.~\ref{eq:poisson} the `simple $\nu$-function', $\nu(x)=0.5+\sqrt{1+4/x}$, is applied, we correct the potential by adding an additional potential $\phi(r) = a_0 r$ to the bulge potential, because this individual $\nu$-function yields the acceleration $g \rightarrow g_\mathrm{N}+a_0$ in the Newtonian limit $a_0 \rightarrow 0$ and therefore not exactly $g_\text{N}$.}
For the disk setup, the full Milgromian gravitational potential of both, the disk and the bulge, is needed. Eq.~\ref{eq:poisson} therefore needs to be solved for the disk+bulge density distribution. This is again performed using the \POR{} code.

Because the critical suface density, $\Sigma_\mathrm{crit}$, looses its meaning in the bulge-dominated area, we adjust the radial velocity dispersion profile to
\begin{equation}
	\sigma_r(R) \propto \exp(-R/(2 R_\mathrm{d})) 
\end{equation}
(e.g., \cite{WD99}).
The proportionality constant is defined by the requirement that $Q(R_0) = Q_0$, with $R_0$ being a constant radius and $Q_0$ a dimensionless constant. This assumption is in agreement with observations \cite{Sohn2012}, and yields a radial velocity dispersion profile similar to Eq.~\ref{eq:sigr1} (apart from the galactic center).

Here, we set up a disk with the parameters similar to those used in Sect.~\ref{sect:disk}, here with $Q(2 R_\mathrm{d}) = 2$, and a Plummer bulge having the mass $M_\mathrm{b} = 0.6\ttt{10}\,\Msun$ and a half-mass radius of $1\unit{kpc}$.

We let this model evolve using the same parameters as for the disk-only model and present the result in Fig.~\ref{fig:diskbulgeevolution}. 
This disk-bulge-model is notably more stable than the bulgeless disk model. In particular, the bulge potential strongly weakens the formation of the galactic bar. This happens to be the case because the bulge itself, as a pressure-supported system, is generally more stable than the rotation-supported disk component, providing the disk with an external potential having a stabilizing effect. Because the bulge makes up only a small fraction of the total mass ($\approx 10\%$), it delays the formation of substructure but can not suppress it.
As a consequence, the non-axisymmetry of the galactic potential grows continuously and more slowly than in the bulge-less model (see Fig.~\ref{fig:barstrength}, dashed line). Also, this model shows an asymmetry peak in this figure due to the formation of spiral arms.

\section{Summary and outlook}

Milgromian dynamics has been very successful on galaxy scales in the last three decades. Most former predictions and tests have however been essentially of static nature, not studying the dynamical evolution of complex self-gravitating systems (with a few exceptions). The small number of efficient $N$-body codes has prevented from testing this theory in as many systems as the standard DM-based model. 

Here, we presented a new Milgromian dynamics $N$-body code, which is a customized version of the \RAMSES{} code. This code handles particles as well as gas dynamics, and provides a grid architecture based on the AMR technique, making it the right tool for a large range of possible applications.
We demonstrated the good performance of the code based on static and dynamical benchmark models for which the analytical solutions are known. Furthermore, we presented models of rotating exponential disk models, with and without a central spherical bulge, thereby showing that we have at hand the tools necessary to set up such dynamical equilibrium models in the Milgromian DM-free framework.

Having such a code at hand, and being able to set up any kind of spherical equilibrium model as well as more complex disk galaxies, we will be able to build more sophisticated dynamical tests by combining these (e.g., galaxy mergers, formation of tidal dwarf galaxies, formation of tidal streams, etc.). Our future goals with this code are a) to test if already observed systems are in agreement or in contradiction with the theory, b) to make predictions that can observationally be tested, and c) to work out general, qualitative differences between Milgromian dynamics and the standard model in view of the dynamical consequences of each of the frameworks.

\section*{Acknowledgements}

The authors thank R.~Teyssier very much for developing and publishing the \RAMSES{} code, and for comments on the manuscript. 
We also thank W.~Dehnen for making his software publicly available in the course of the \NEMO{} package. 
FL is partly supported by DFG grant \mbox{KR1635/16-1} and by a stipend from the rectorate of the University of Bonn. 
The presented code will be made publicly available in the future.


\end{document}